\documentclass[aps,prl,showpacs,twocolumn]{revtex4}

\usepackage[utf8x]{inputenc}
\usepackage{amsmath}
\usepackage{amssymb}
\usepackage{amsfonts}
\usepackage{graphics}
\usepackage[pdftex]{graphicx}
\usepackage[normal]{subfigure}
\usepackage{color}
\begin{document}

\title{Quantum walks as massless Dirac Fermions in curved Space-Time}
\author{Giuseppe Di Molfetta$\dagger$}
\author{M. Brachet*}
\author{Fabrice Debbasch$\dagger$} 
\affiliation{*CNRS, Laboratoire de Physique Statistique 
Ecole Normale Sup\'erieure, 
75231 Paris Cedex 05, France\\
$\dagger$ LERMA, UMR 8112, UPMC and Observatoire de Paris, 61 Avenue de l'Observatoire  75014 Paris, France}

\date{\today}

\begin{abstract}

A particular family of time- and space-dependent discrete-time quantum walks (QWs) is considered in one dimensional physical space. The continuous limit of these walks is defined through a new procedure and computed in full detail. In this limit, the walks coincide with the propagation of a massless Dirac fermion in an arbitrary gravitational field. A QW mimicking the radial propagation of a fermion outside and inside the event horizon of a Schwarzschild black hole is explicitly constructed and simulated numerically. 
Thus, the family of QWs considered in our manuscript provides a new analogue system to study experimentally coherent quantum propagation in curved space-time.

\end{abstract}

\maketitle

The first quantum walk (QW) was built  by Feynman \cite{Schweber86a} as a possible discretization of the standard, massive Dirac dynamics in flat space-time. General discrete-time QWs have then been introduced in the physics literature by \cite{ADZ93a} and \cite{Meyer96a} and the continuous-time version first appeared in \cite{FG98a}. QWs are the simplest formal analogues of classical random walks and are important in many fields, ranging from fundamental quantum physics \cite{var96a,Perets08a} to quantum algorithmics \cite{Amb07a,MNRS07a}, solid state phsyics \cite{Bose03a,Aslangul05a,Burg06a, Bose07a} and biophysics \cite{Engel07a,Collini10a}. 

QWs have been realized experimentally, for example as transport of  trapped ions \cite{Schmitz09a,Zahring10a}, of photons in wave guide lattices \cite{Perets08a} or optical networks \cite{Schreiber10a} and of atoms in optical lattices \cite{Karski09a}. QW experiments of two photons \cite{Peruzzo10a} have recently been performed, with the possibility of simulating Bose or Fermi statistics \cite{Sansoni11a}, and cavity QED QWs have also been proposed \cite{Sanders03a}.

Following Feynman's idea, several authors have studied the continuous limit of general QWs. Most publications \cite{FeynHibbs65a,KRS03a,BH04a,Strauch06a,Strauch06b,Strauch07a,BES07a,Chandra10a} only envisage QWs with constant coefficients. The continuous limit of QWs with time-and space-dependent coefficients has been considered only recently, in \cite{DMD12a,DDMEF12a,DDM13a}. These references present several families of  QWs, in both $(1 + 1)$ or $(1+ 2)$ space-time dimensions, whose continuous limit is described by a flat space-time Dirac equation with generalized mass term and electromagnetic coupling. The electromagnetic field 
is generated by the space-time dependence of the angles defining the walks and thus vanishes if these  angles are constant.

The other gauge field wich couples naturally to a Dirac spinor is evidently gravity. Yet, QWs whose continuous limit are described by Dirac equations in curved space-time have remained elusive. This article considers a particular family of discrete-time QWs with non constant angles in $(1 + 1)$ space-time dimensions and associates to this family a continuous limit which is described by a massless Dirac equation in curved space-time. 
This family has not been previously investigated in \cite{DMD12a,DDMEF12a,DDM13a} and the limit procedure used in these references cannot be applied to this new family.
The result presented in this article opens the way to possible laboratory experiments simulating coherent quantum propagation in relativistic gravitational fields. It also establishes a connection between general relativity and all the afore mentioned fields, where QWs are useful.

We consider QWs defined over discrete time and discrete one dimensional space, driven by
time- and space-dependent quantum coins acting on a two-dimensional Hilbert space $\mathcal H$. 
The walks are defined by the following finite difference equations, valid for all $(j, m) 
\in \mathbb{N}  \times \mathbb{Z}$: 
\begin{equation}
\begin{bmatrix} \psi^{L}_{j+1, m }\\ \psi^{R}_{j+1, m } \end{bmatrix} \  = 
B\left( \theta_{j, m} \right)
 \begin{bmatrix} \psi^{L}_{j, m+1} \\ \psi^{R}_{j, m-1} \end{bmatrix},
\label{eq:defwalkdiscr}
\end{equation}
where 
\begin{equation}
 B(\theta) = 
\begin{bmatrix} - \cos\theta & i \sin\theta\\ - i \sin\theta &  + \cos\theta
 \end{bmatrix}
\label{eq:defB}
\end{equation} 
The index $j$ labels instants  and the index $m$ labels spatial points.
For each instant $j$ and each spatial point $m$, the wave function $\Psi_{jm} = \psi^L_{jm} b_L + \psi^R_{jm} b_R$ has two components $\psi^L_{jm} $ and $\psi^R_{jm}$ on the spin basis $(b_L, b_R)$ and these code for the probability amplitudes 
of the particle jumping towards the left or towards the right. Note that the spin basis is interpreted as being independent of $j$ and $m$.
The total probability $\pi_j= \sum_m \left(
\mid \psi^L _{j, m} \mid ^2+ \mid \psi^R _{j, m}\mid^2 \right)$ is independent of $j$ {\sl i.e.} 
conserved by the walk.
The set of angles $\left\{ \theta_{j, m}, (j, m) \in \mathbb{N} 
 \times \mathbb{Z}\right\}$ 
defines the walks and is arbitrary.

Consider now, for all $(n, j) \in {\mathbb N}^2$, the collection 
$W_j^n = (\Psi_{k, m})_{k = nj, m \in \mathbb Z}$.
This collection represents the state of the QW at `time' $k = nj$. For any given $n$, the collection $S^n = (W_j^n)_{j \in \mathbb N}$ thus represents the entire history of the QW observed through a stroboscope of `period' $n$. The evolution equations for $S^n$ are those linking $W_{j+1}^n$ to $W_j^n$ for all $j$. These can be deduced from the original evolution equations (\ref{eq:defwalkdiscr}) of the walk, which also coincide with the evolution equations of $S^1$. In particular, the evolution equations of $S^2$ read:
\begin{eqnarray}
\hspace{-0.4cm}
\psi^L_{j+2, m} & = & c_{j+1, m} \left( c_{j, m+1} \psi^L_{j, m+2} - i s_{j, m+1} \psi^R_{j, m}\right) \nonumber\\
& & +  s_{j+1, m} \left( s_{j, m-1} \psi^L_{j, m} + i c_{j, m-1} \psi^R_{j, m-2}\right), \label{eq:walk1}
\end{eqnarray}
\begin{eqnarray}
\hspace{-0.5cm}
\psi^R_{j+2, m} & = & s_{j+1, m} \left( i c_{j, m+1} \psi^L_{j, m+2} + s_{j, m+1} \psi^R_{j, m}\right) \nonumber\\
& & -  c_{j+1, m} \left( i s_{j, m-1} \psi^L_{j, m} - c_{j, m-1} \psi^R_{j, m-2}\right),\label{eq:walk2}
\end{eqnarray}
where $c_{jm} = \cos(\theta_{jm})$ and  $s_{jm} = \sin(\theta_{jm})$.

To investigate the continuous limit of $S^n$,
we first introduce a time step $\Delta t$ and a space step $\Delta x$.
We then consider that $\Psi_{jm}$ and $\theta_{jm}$ are the values taken 
by a two-component wave function $\Psi$ and by a function $\theta$ at the space-time point
$(t_j = j \Delta t, x_m = m \Delta x)$.
We finally suppose, that $\Psi$ and $\theta$ are at least twice differentiable with respect to both space and time variables for all 
sufficiently small values of $\Delta t$ and $\Delta x$. 
The formal continuous limit of $S^n$ is defined as the couple of differential equations obtained from the discrete-time evolution equations defining $S^n$ 
 by letting both $\Delta t$ and $\Delta x$ tend to zero.  
Let us therefore introduce a time-scale ${\mathcal T}$, a length-scale ${\mathcal L}$, an infinitesimal $\epsilon$ and write $\Delta t = \epsilon {\mathcal T} $ and $\Delta x = \epsilon {\mathcal L}$. The continuous limit of $S^n$ can then be investigated by Taylor expanding in powers of $\epsilon$ the discrete equations defining $S^n$. For the limit to exist, all zeroth order terms must identically cancel each other and the differential equation describing the imit is then obtained by equating to zero the non identically vanishing, lowest order contribution. Consider first $S^1$, which is identical to the original walk. It is rather obvious that zeroth order terms cancel each other 
only if the operator $B$ defining the walk tends to unity as $\epsilon$ tends to $0$ (see \cite{DMD12a} for a detailled discussion of this point). The operator $B$ defining the family of walks considered in this Letter does not depend on $\epsilon$ and is different from unity for all values of $\theta$.
Thus, $S^1$ does not admit a continuous limit for the family of walks defined by (\ref{eq:defwalkdiscr}). But $S^2$ on the other hand does. Indeed a straightforward computation 
delivers the following 
equation obeyed by the wave function $\Psi$:
\begin{equation}
\Psi_T + (\cos \theta) P \Psi_X = Q \Psi,
\label{eq:DiracMR}
\end{equation}
where the operators $P$ and $Q$ are represented, in the base $(b_L, b_R)$, by the matrices:
\begin{equation}
P = \left(
\begin{matrix}
- \cos \theta & i \sin \theta \\ - i \sin \theta & \cos \theta
\end{matrix}
\right)
\end{equation}
and
\begin{equation}
Q = \left(
\begin{matrix}
- \theta_X\, \frac{\sin 2 \theta}{2} &  \frac{i}{2} \left( \theta_T - \theta_X (\cos 2 \theta)\right) \\ \frac{i}{2} \left( \theta_T + \theta_X (\cos 2 \theta)\right) & \theta_X\, \frac{\sin 2 \theta}{2}
\end{matrix}
\right).
\label{eq:defQ}
\end{equation}
In (\ref{eq:DiracMR}) and (\ref{eq:defQ}), the subscript $T$ ({\sl resp.} $X$) indicates a derivative with respect to the dimensionless variable $T = t/ {\mathcal T}$ ({\sl resp.} $X = x/{\mathcal L}$).

The operator $P$ is self-adjoint and its eigenvalues are $-1$ and $+1$. Two
eigenvectors associated to these eigenvalues are 
\begin{equation}
b_- = i \left(\cos \frac{\theta}{2} \right) b_L - \left(\sin \frac{\theta}{2} \right) b_R,
\end{equation}
\begin{equation}
b_+ = i \left(\sin \frac{\theta}{2}\right) b_L + \left( \cos \frac{\theta}{2}\right) b_R.
\end{equation}
The family $(b_-, b_+)$
forms an orthonormal basis of the two dimensional spin Hilbert space.
Let us now rewrite equation (\ref{eq:DiracMR}) in components, but in this new orthonormal basis.
A tedious but straightforward computation leads to:
\begin{eqnarray}
\psi^-_T - (\cos \theta) \psi^-_X + \frac{\theta_X}{2}\, (\sin \theta)\,  \psi^- & = & 0 \nonumber \\
\psi^+_T + (\cos \theta) \psi^+_X - \frac{\theta_X}{2}\, (\sin \theta)\,  \psi^+ & = & 0,
\label{eq:definitive}
\end{eqnarray}
where $\psi^-$ and $\psi^-$ are the components of $\Psi$ in the new basis. This form of the equations makes it easy to check that the continuous dynamics conserves the total probability
$\pi (T) = \int dX \mid \Psi (T, X) \mid^2 = \int dX \left(\mid \psi^- (T, X) \mid^2 + \mid \psi^+ (T, X) \mid^2 \right)$, as it should.

Suppose now, to make the discussion definite, that $\cos \theta$ is strictly positive and introduce in space-time $ \left\{(T, X) \right\}$ the Lorentzian, possibly curved metric $G$ defined by its covariant
components $(G_{\mu \nu}) = \mbox{diag}(1, - 1/\cos^{2} \theta)$,
where $(\mu, \nu) \in \left\{T, X \right\}^2$. 
This metric defines the canonical, scalar `volume' element 
${\mathcal D}_G X = \sqrt{- G}\,  dX = {dX}/{\cos \theta}$ in physical $1$D $X$-space, where $G$ is the determinant of the metric components $G_{\mu \nu}$.
Dirac spinors are normalized to unity with respect to ${\mathcal D}_G X$, whereas $\Psi$ is normalized to unity with respect to $dX$. We thus introduce $\Phi = \Psi \sqrt{\cos \theta}$ and rewrite the equations of motion (\ref{eq:definitive}) in terms of $\Phi$. We obtain:
\begin{eqnarray}
\gamma^a\left[ 
e^\mu_a \partial_\mu \Phi 
+ \, \frac{1}{2}\, 
\frac{1}{\sqrt{-G}}\,  \partial_\mu \left(
\sqrt{-G} e^\mu_a
\right) \Phi
\right] = 0,
\label{eq:DIRAC}
\end{eqnarray}
where $\mu \in \left\{T, X\right\}$, $a \in \left\{0, 1\right\}$. The usual $2D$ gamma matrices are:
\begin{equation}
\gamma^0= 
\left(
\begin{matrix}
0 & 1\\ 1 & 0
\end{matrix}
\right), \, \,
\gamma^1 = 
\left(
\begin{matrix}
0 & 1\\ - 1 & 0
\end{matrix}
\right)
\end{equation}
and the $e^\mu_a$ are the components of the diad (orthonormal basis)  
$e_0 = e_T$ and $e_1 = (\cos \theta)\,  e_X$ on the original coordinate basis $(e_T, e_X)$. 
Equation (\ref{eq:DIRAC}) is the standard \cite{SR94a} equation of motion for a massless Dirac spinor in $(1 + 1)$ dimensional space-time with metric $G$. The spin basis is $(b_-, b_+)$.

This result shows that QWs can be used to model quantum transport in any $2D$ gravitational field. Indeed, 
any $2D$ Lorentzian metric can be put under the above diagonal form 
by a suitable choice of coordinates. The single angle  $\theta(t, x)$ is thus enough to describe any $2D$ gravitational field. Let us stress however that
gravity is very different in $2D$ and in $4D$ since, In particular, all $2D$ space-times are conformally flat \cite{Wald84a}. 
But equation (\ref{eq:DIRAC}) can also be used to model quantum transport in higher dimensional space-times by QWs on the line. As an example, we now construct a QW on the line which mimicks the radial motion of a Dirac spinor in a spherically symmetric $4D$ black hole.

A Schwarschild black hole is a spherically symmetric solution of Einstein equation in vacuo. 
The corresponding $4D$ metric reads, in dimensionless Lema\^itre coordinates $(\tau, \rho, \theta, \phi)$ \cite{L33a}:
\begin{equation}
ds^2 = d\tau^2- \frac{r_g}{r}\,  d\rho^2 - r^2 d\Omega
\end{equation}
where 
$r (\tau, \rho) = r_g^{1/3} \left[ \frac{3}{2}\left(\rho - \tau\right)\right]^{2/3}$,
$d\Omega = d\theta^2 + (\sin^2 \theta) d\phi^2$. The event horizon is located at $r = r_g$ {\sl i.e.} $\rho = \tau + (2/3) r_g$, and the singularity is located at $r = 0$ {\sl i.e.} $\rho = \tau$. The exterior of the black hole is the domain $r > r_g$.The range of variations for the Lema\^itre coordinates is $\tau \ge 0$, $\rho \ge \tau$ ({\sl i.e.} $r (\tau, \rho) \ge 0$), $0 \le \theta \le \pi$, $0 \le \phi < 2 \pi$.

Because of the spherical symmetry, a point mass which starts its motion radially will go on moving radially. Radial motion can be studied by introducing the $2D$ metric $g$, also singular at $r=0$, with covariant components $g_{\tau \tau} = 1$, $g_{\rho \rho} = - r_g/r$, $g_{\tau \rho} = g_{\rho \tau} = 0$.
The null geodesics of $g$ are defined by $d \tau = \pm \left( r_g/r(\tau, \rho)\right)^{1/2} d\rho$. Note that the $2D$ projection of the horizon on the $(\tau, \rho)$-plane coincides with a null geodesics of $g$.

\begin{figure}[h!]
\includegraphics[width=9cm]{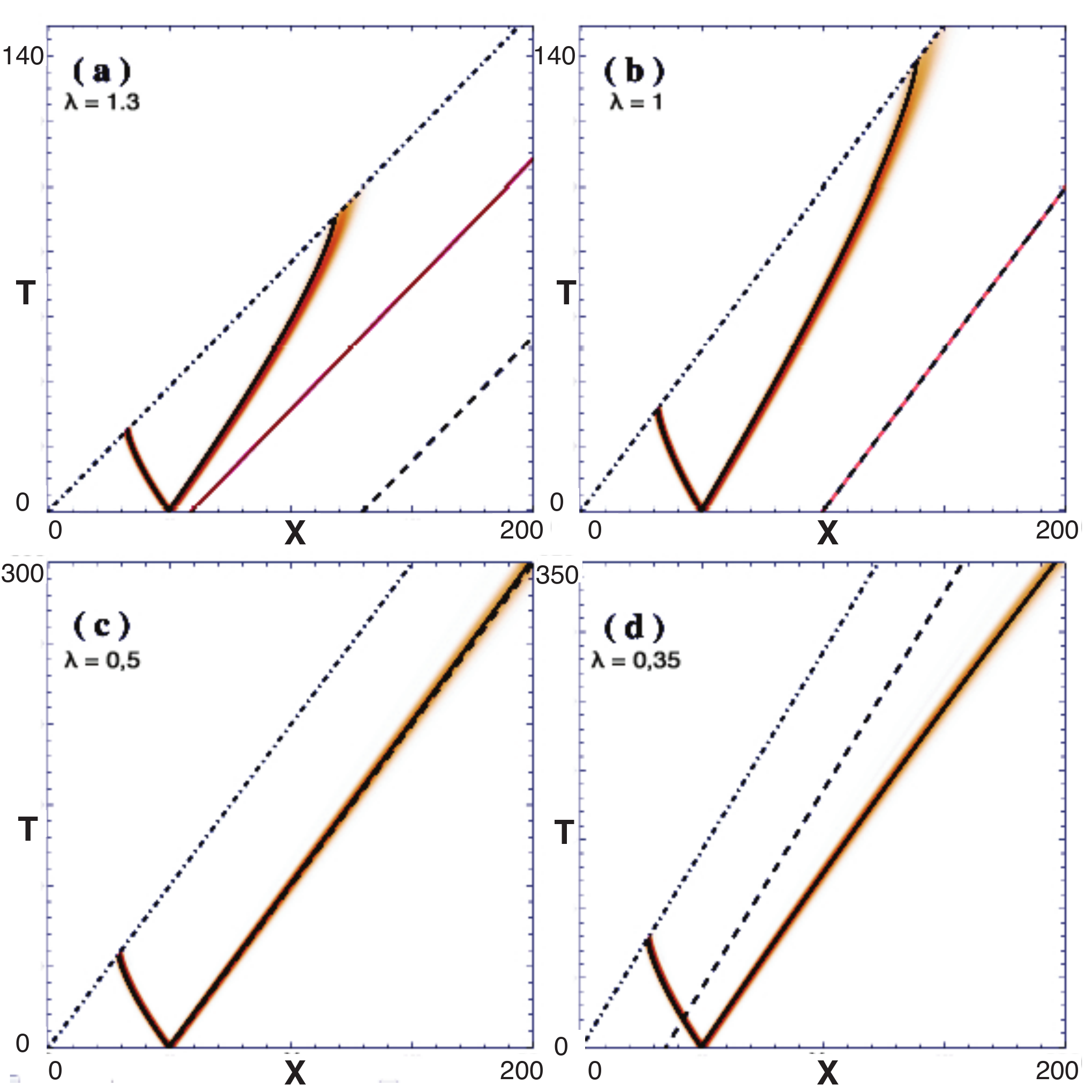}
 \caption{(color online) Density of the QW vs. null geodesics (solid curves) of the $2D$ Schwarschild metric for various values of $\lambda$ (see text and Eq.(\ref{eq:defr})). The singularity is represented by the dotted and dashed line on the left and the horizon
(which is a null geodesic) is represented by the dashed line. Figure (a) and (b): The two branches of the QW which starts inside the horizon end up on the singularity. The (red) solid line represents the limit of the definition domain ${\mathcal D}$ of the QW. Figure (c): one branch of a QW which starts on the horizon stays on the horizon while the other branch ends up on the singularity. Figure (d): one branch of  the QW which starts outside the horizon propagates away from the black hole, and the other branch ends up on the singularity}.
 \label{fig:QW}
  \end{figure}

We now identify the dimensionless time $T$ with the time coordinate $\tau$ and the dimensionless space variable $X$ with $\lambda \rho$, where 
$\lambda$ is an arbitrary strictly positive real number (see Fig.\ref{fig:QW}). The `radius' $r$ can then be expressed as a function of $T$ and $X$:
\begin{equation}
r (T, X) = \left[\frac{3}{2}\left(\frac{X}{\lambda} - T\right)\right]^{2/3}r_g^{1/3},
\label{eq:defr}
\end{equation}
and the components of $g$ in the coordinate basis associated to $T$ and $X$ are
$g_{TT} = 1$, $g_{XX} = - r_g/(\lambda^2 r)$, $g_{TX} = g_{XT} = 0$. 
Note that the condition $\rho \ge \tau$ transcribes into $X \ge \lambda T$. 

Let $\mathcal D$ be the domain where $-g_{XX} \ge1$. This domain is characterized, in $(T, X)$ coordinates, by the condition
$X \leq \lambda T + \frac{2}{3 \lambda^2}\, {r_g}$.
In $\mathcal D$ the metric $g$ can be identified with the metric $G$ (see Eq.(\ref{eq:metric})).
This identification defines a angle $\theta$ which depends on $T$ and $X$ by: 
\begin{equation}
(\cos \theta) (T, X) = \lambda\, \sqrt{\frac{r(T, X)}{r_g}}
\label{eq:deftheta}
\end{equation}
and, by extension, a QW in $\mathcal D$. 

The condition defining $\mathcal D$ can be rewritten as $r \le r_g/\lambda^2$, The domain $\mathcal D$ thus includes, for all $\lambda$, the singularity located at $r = 0$. For $\lambda >1$, $r_g/\lambda^2 < r_g$ and $\mathcal D$ is then entirely located inside the horizon.
For $\lambda = 1$, $\mathcal D$ coincides with the interior of the horizon, and $\mathcal D$ extends outside the horizon for $\lambda < 1$. 

The spatial density $\mid \Psi(T, X) \mid^2$ of the walk defined by Eqs.(\ref{eq:defwalkdiscr}),(\ref{eq:defB}) and (\ref{eq:deftheta}) is plotted in Fig.\ref{fig:QW}. All graphs have been obtained with $r_g = 150$ and $\epsilon = 0.5$. The initial condition is $\Psi(0, X) = \sqrt{n_0(X)} \left( b_L + i b_R \right)$ with an initial Gaussian density $n_0$
of variance $\Delta X_0 = 2.5$ centered on $X_0 = 50.5$.
In Fig.\ref{fig:QW}.a and Fig.\ref{fig:QW}.b (in which the right limit of $\mathcal D$ coincides with the horizon) both branches of the QW starts their evolution inside the horizon, and end up on the singularity. In Fig.\ref{fig:QW}.c the QW starts exactly on the horizon; the right branch follows it while the left branch ends up on the singularity. In Fig.\ref{fig:QW}.d the QW starts outside the horizon; its right branch propagates away from the horizon while the left branch still ends up on the singularity. 

Fig.\ref{fig:QW} clearly shows that the QW closely follows null geodesics of the metric {\sl i.e.} behaves as a massless fermion in the gravitational field of the black hole. The agreement between the geodesics and the density profile of the walk is all the more remarkable given that all graphs correspond to $\epsilon = 0.5$, which lies well outside of the continuous limit $\epsilon \ll 1$ envisaged above. Note however that the right branch of the QW lags slightly behind the null geodesic when approaching the $r=0$ singularity (see top of Fig.\ref{fig:QW}.b) The numerical results thus suggest that the main result of this Letter can somehow be extended beyond the continuous limit. This interesting point will be investigated in future publications.

In summary, we have considered a particular family of one-dimensional QWs 
which does not admit a continuous limit by the procedure used in \cite{DMD12a,DDMEF12a,DDM13a}.
We have introduced a new procedure, which ammounts to keeping one time step out of two, and which gives a continuous limit to walks from this family.
We have computed this limit and proven that it coincides with the propagation of a massless Dirac fermion in an arbitrary gravitational field. Note however that Fig.\ref{fig:QW}, where all time steps are retained, shows that the density clearly follows, at all times, the null geodesic predicted by the continuous limit. We have also constructed explicitely a QW which mimicks the propagation of a fermion outside and inside the event horizon of a Schwarzschild black hole, and illustrated this construction by numerical simulations. 

Let us now discuss the above results.
Keeping only one time-step of the QW out of two to build a continuous limit might appear unnatural and might even look like a purely mathematical trick. Let us explain now in detail why this is not so. Consider first, as an instructive example, the sequence of numbers $u_j$ defined by $u_0 = 1$ and
\begin{equation}
u_{j+1} = \sigma \exp(i \omega {\mathcal T}) u_j, 
\label{eq:defu}
\end{equation}
for $j \in \mathbb N$; here, $\omega$ is an arbitrary real number, $\mathcal T$ is a time-scale
and $\sigma = \pm 1$ and does not depend on $j$. A direct computation shows that $u_j = \sigma^j \exp(i \omega t_j)$ with $t_j = j \mathcal T$.  Suppose $\sigma = +1$. The sequence $u_j$ is then a simple circular function of the time $t_j$. On the contrary, if $\sigma = -1$, the sequence $u_j$ is then  circular function of the time $t_j$ combined with an extra phase shift of $\pi$ at every time-step. The sequence $u_j$ admits a continuous limit if $\sigma = +1$. But, it does not if $\sigma = -1$, because of this extra phase shift.
In particular, if $\sigma = -1$, the oscillating behavior of the sequence $u_j$ cannot be recovered by simply taking the continuous limit of the evolution equation (\ref{eq:defu}). The best way to recover this oscillating behavior is then to consider the new sequence $v_k$ built out of $u_j$ by keeping only one time step out of two. Indeed, this new sequence obeys the discrete evolution equation $v_{k+1} = \exp(2 i \omega {\mathcal T}) v_k$. This equation admits a continuous limit, described by the ordinary differential equation $\frac{dv}{dt} = 2 i \omega\, v$, which clearly reveals the oscillating behaviour in $v$ and, thus, in $u$. Naturally, all information on the extra phase-shift of $\pi$ is lost in this procedure. Note however that this extra phase-shift does not influence the sequence of the squared moduli $\mid u_j \mid^2$.

Let us now compare the preceding example with the QWs examined in \cite{DMD12a,DDMEF12a,DDM13a} and in this article. The spinor $\Psi$ plays a role similar to $u$ and the evolution equation of the walk (equation (\ref{eq:defwalkdiscr}) for the QWs considered in this article) has the same status as (\ref{eq:defu}). The QWs studied in \cite{DMD12a,DDMEF12a,DDM13a} are equivalent to the sequence $u_j$ obtained with $\sigma = +1$. They thus admit a standard continuous limit, which is fully described in these earlier publications. This limit coincides with the propagation in flat space-time of a Dirac fermion possibly coupled to an electric field. On the other hand, the QWs 
defined by (\ref{eq:defwalkdiscr}) and (\ref{eq:defB})
correspond to the sequence $u_j$ with $\sigma = -1$ and they do not admit a continuous limit. But the derived walks built by keeping only one time-step out of two of the original QWs   
do admit a continuous limit because the squared matrix ${\mathcal B}^2 = 1$, just as $(-1)^2 = 1$. Keeping one time-step out of two is thus not a contrived, unphysical procedure. On the contrary, it is 
dictated by the very definition of the QWs we study in this article and it is the only one which delivers a continuous limit for these walks. 
The obtained continuous limit coincides with the physically interesting situation of a massless Dirac fermion propagating in curved space-time.
Since the retained limit procedure is itself dictated by the QWs studied in this article, the geometry appearing in their continuous limit 
is an intrinsic property of the walks themselves.


The explicit construction of a QW mimicking propagation in and around a black hole shows that it is possible, at least in principle, to simulate by laboratory experiments the propagation of quantum systems in interesting relativistic gravitational fields.

The work presented in this article should naturally be extended in several directions. One should first investigate systematically all QWs on the line defined with a quantum coin acting on a $2$D Hilbert space and, for each family of walks, 
try and determine if and how a continuous limit can be obtained.
One should also extend the main result of this article to QWs defined on physical space of higher dimension and/or defined by quantum coins acting on a higher dimensional Hilbert space. A particular goal would be to obtain walks whose limits are described by a Dirac fermion coupled to both a gravitational and an electromagnetic field.

As noted earlier, the geometry appearing in the continuous limit of the QWs studied in this article is an intrinsic property of these walks and
is not to be confused with
the geometry the `real', `physical' space-time which might be used to realize the walk experimentally. Another extension of this work would therefore be to consider walks defined on curved physical spaces ({\sl resp.} graphs), and to investigate how the geometry of the underlying space ({\sl resp.} graph) couples to the intrinsic geometry of the walk. This is not a purely academical problem, since QWs can model photon transport in networks of algae, which may have a non trivial geometry. 

The main result of this Letter also suggests that concepts from general relativity and differential geometry may play a key role to understanding the behaviour of QWs on graphs and their use in quatum algorithmics. This role should also be investigated thoroughly. 

\def\cprime{$'$} \def\cprime{$'$} \def\cprime{$'$} \def\cprime{$'$}
  \def\cprime{$'$}


\begin{thebibliography}{10}

\bibitem{Schweber86a}
S.S. Schweber.
\newblock Feynman and the visualization of space-time processes.
\newblock {\em Rev. Mod. Phys.}, 58:449, 1986.

\bibitem{ADZ93a}
Y.~Aharonov, L.~Davidovich, and N.~Zagury.
\newblock Quantum random walks.     
\newblock {\em Phys. {R}ev. A}, 48:1687, 1993.

\bibitem{Meyer96a}
D.A. Meyer.
\newblock Quantum mechanics of lattice gas automata i. one particle plane waves
  and potentials.
\newblock {\em {J}. {S}tat. {P}hys.}, 85:551, 1996.

\bibitem{FG98a}
E.~Farhi and S.~Gutmann.
\newblock Quantum computation and decision trees.
\newblock {\em Phys. {R}ev. A}, 58:915, 1998.

\bibitem{var96a}
D.~Giulini, E.~Joos, C.~Kiefer, J.~Kupsch, I.-O. Stamatescu, and H.D. Zeh.
\newblock {\em Decoherence and the appearance of a Classical World in Quantum
  Theory}.
\newblock Springer-Verlag, Berlin, 1996.

\bibitem{Perets08a}
H.B. Perets, Y.~Lahini, F.~Pozzi, M.~Sorel, R.~Morandotti, and Y.~Silberberg.
\newblock Realization of quantum walks with negligible decoherence in waveguide
  lattices.
\newblock {\em Phys. Rev. Lett.}, 100:170506, 2008.

\bibitem{Amb07a}
A.~Ambainis.
\newblock Quantum walk algorithm for element distinctness.
\newblock {\em SIAM Journal on Computing}, 37(1):210--239, 2007.

\bibitem{MNRS07a}
F.~Magniez, A.~Nayak, J.~Roland, and M.~Santha.
\newblock Search via quantum walk.
\newblock In {\em Proceedings of the thirty-ninth annual ACM symposium on
  Theory of computing}, New {Y}ork, 2007. ACM.

\bibitem{Aslangul05a}
C.~Aslangul.
\newblock Quantum dynamics of a particle with a spin-dependent velocity.
\newblock {\em Journal of Physics A: Mathematical and Theoretical},
  38(1):1--16, 2005.

\bibitem{Bose03a}
S.~Bose.
\newblock Quantum communication through an unmodulated spin chain.
\newblock {\em Phys. Rev. Lett.}, 91:207901, 2003.

\bibitem{Burg06a}
D.~Burgarth.
\newblock {\em Quantum state transfer with spin chains}.
\newblock PhD thesis, University College London, 2006.

\bibitem{Bose07a}
S.~Bose.
\newblock {\em Contemp. {P}hys.}, 48:13, 2007.

\bibitem{Collini10a}
E.~Collini, C.Y. Wong, K.E. Wilk, P.M.G. Curmi, P.~Brumer, and G.D. Scholes.
\newblock {\em Nature}, 463:644, 2010.

\bibitem{Engel07a}
G.S. Engel, T.R. Calhoun, R.L. Read, T.-K. Ahn, T.~Manal, Y.-C. Cheng, R.E.
  Blankenship, and G.~R. Fleming.
\newblock {\em Nature}, 446:782, 2007.

\bibitem{Schmitz09a}
H.~Schmitz et~al.
\newblock {\em Phys. Rev. Lett.}, 103:090504, 2009.

\bibitem{Zahring10a}
F.~Z\"ahringer, G.~Kirchmair, R.~Gerritsma, E.~Solano, R.~Blatt, and C.F. Roos.
\newblock Realization of a quantum walk with one and two trapped ions.
\newblock {\em Phys. Rev. Lett.}, 104:100503, 2010.

\bibitem{Schreiber10a}
A.~Schreiber et~al.
\newblock Photons walking the line.
\newblock {\em Phys. Rev. Lett.}, 104:050502, 2010.

\bibitem{Karski09a}
M.~Karski et~al.
\newblock {\em Science}, 325:174, 2009.

\bibitem{Peruzzo10a}
A.~Peruzzo et~al.
\newblock {\em Science}, 329(5998):1500, 2010.

\bibitem{Sansoni11a}
L.~Sansoni et~al.
\newblock Two-particle bosonic-fermionic quantum walk via 3d integrated
photonics.
\newblock {\em Phys. Rev. Lett.}, 108: 010502, 2012. 


\bibitem{Sanders03a}
B.C. Sanders, S.D. Bartlett, B.~Tregenna, and P.L. Knight.
\newblock {\em Phys. {R}ev. A}, 67:042305, 2003.

\bibitem{BES07a}
A.J.~Bracken, D.~Ellinas and I.~Smymakis.
\newblock {Free-Dirac-particle evolution as a quantum random walk}.
\newblock {\em Phys. Rev. A}, 75:022322, 2007.

\bibitem{BH04a}
Ph. Blanchard and M.-O. Hongler.
\newblock Quantum random walks and piecewise deterministic evolutions.
\newblock {\em Phys. Rev. Lett.}, 92(12):120601--1--120601--4, 2004.

\bibitem{Chandra10a}
C.M. Chandrasekhar, S.~Banerjee, and R.~Srikanth.
\newblock Relationship between quantum walks and relativistic quantum
  mechanics.
\newblock {\em Phys. {R}ev. A}, 81:062340, 2010.

\bibitem{FeynHibbs65a}
R.P. Feynman and A.R. Hibbs.
\newblock {\em Quantum Mechanics and Path Integrals}.
\newblock International Series in Pure and Applied Physics. McGraw-Hill Book
  Company, 1965.

\bibitem{KRS03a}
P.L. Knight, E.~Rold\`an, and J.E. Sipe.
\newblock Quantum walk on the line as an interference phenomenon.
\newblock {\em Phys. {R}ev. A}, 68:020301, 2003.

\bibitem{Strauch06a}
F.W. Strauch.
\newblock Relativistic quantum walks.
\newblock {\em Phys. {R}ev. A}, 73:054302, 2006.

\bibitem{Strauch06b}
F.W. Strauch.
\newblock Connecting the discrete- and continuous-time quantum walks.
\newblock {\em Phys. {R}ev. A}, 74:030301, 2006.

\bibitem{Strauch07a}
F.W. Strauch.
\newblock Relativistic effects and rigorous limits for discrete-time and continuous-time quantum walks.
\newblock {\em J. {M}ath. {P}hys.}, 48:082102, 2007.

\bibitem{DMD12a}
G. Di Molfetta and F. Debbasch.
\newblock Discrete-time quantum walks: Continuous limit and symmetries.
\newblock {\em J.Math.Phys.}, 53:123302, 2012.

\bibitem{DDM13a}
F.~Debbasch, G. Di Molfetta.
\newblock Discrete-time quantum walks: Continuous limit in 1 + 1 and 1 + 2 dimension.
\newblock {\em J.Comp.Th.Nanosc.}, {\em to be published}, 2013. 

\bibitem{DDMEF12a}
F.~Debbasch, G. Di Molfetta, D.~Espaze, and V. Foulonneau.
\newblock Propagation in quantum walks and relativistic diffusions.
\newblock {\em Phys. Scr.}, T 151:014044, 2012.

\bibitem{SR94a}
A. Sinha and R. Roychoudhury.
\newblock Dirac equation in (1 + 1)-dimensional curved space-time.
\newblock{\em Int.J.Th.Phys.}, 33(7):1511--1522, 1994.

\bibitem{Wald84a}
R. Wald.
\newblock Gravitation.
\newblock{\em University of Chicago Press}, 1984.

\bibitem{L33a}
G. Lema{\^\i}tre.
\newblock L'univers en expansion.
\newblock {\em Ann. Soc. Sc. Bruxelles A}, 53:51--85, 1933.

\end{thebibliography}
\end{document}